\documentstyle[aps,prl,psfig]{revtex}
\newcommand\beq{\begin{equation}}
\newcommand\eeq{\end{equation}}
\newcommand\bea{\begin{eqnarray}}
\newcommand\eea{\end{eqnarray}}

\newcommand\be{\beta}
\newcommand\half{{\frac 12}}
\newcommand\bfeta{\mbox{\boldmath$\eta$}}
\newcommand\bfA{{\mathbf A}}

\newcommand\FF{{\cal F}}
\newcommand\ltwid{\mathrel{\raise.3ex\hbox{$<$\kern-.75em
\lower1ex\hbox{$\sim$}}}}
\newcommand\Journal[4]{{#1} {\bf #2}, #3 (#4)}
\newcommand\prevl{{\em Phys. Rev. Lett.}}
\newcommand\prevb{{\em Phys. Rev. B}}

\begin{document}
\twocolumn[\hsize\textwidth\columnwidth\hsize\csname@twocolumnfalse%
\endcsname


\begin{flushright}
\vskip-0.75cm
UdeM-GPP-TH-01-90\\
McGill 01-15\\
\end{flushright}
\title{Vortex lattice stability in the SO(5) model}

\author{M.\ Juneau, R.\ MacKenzie, M.-A.\ Vachon}
\address{\it Laboratoire Ren\'e-J.-A.-L\'evesque, Universit\'e de
Montr\'eal,
C.P. 6128, Succ. Centre-ville, Montr\'eal, QC H4V 2A5}

\author{J.M.~Cline}

\address{Physics Department, McGill University,
3600 University Street, Montr\'eal, Qu\'ebec, Canada H3A 2T8}

\maketitle

\begin{abstract} 

We study the energetics of superconducting vortices in the SO(5) model for
high-$T_c$ materials proposed by Zhang.   We show that for a wide range of
parameters normally corresponding to type II superconductivity,  the free
energy per unit flux $\FF(m)$ of a vortex with $m$ flux quanta  is a
decreasing function of $m$, provided the doping is close to its critical
value. This implies that the Abrikosov lattice is unstable, a behaviour
typical of type I superconductors.  For dopings far from the critical
value, $\FF(m)$ can become very flat, indicating a less rigid vortex
lattice, which would melt at a lower temperature than  expected for a BCS
superconductor.
\leftline{PACS: 74.20.De, 74.20.Mn, 74.25.Ha, 74.72-h}\\
\end{abstract}

]

\noindent {\bf 1. Introduction.}
The phase diagrams of all high-temperature superconductors have a
rich structure, with two prominent features at low temperature:
antiferromagnetism and superconductivity. Antiferromagnetism (AF) is
seen at low doping, while superconductivity (SC) is observed if the
doping exceeds a critical value.

A description of these phenomena was proposed by S.C. Zhang
\cite{zhang}, who observed that both superconductivity and
antiferromagnetism involve spontaneous symmetry breaking.
Borrowing heavily on ideas from particle physics,
he suggested that the symmetries involved are unified into a larger
approximate symmetry group. He presented a strong case for the group
SO(5), with the SC and AF order parameters combined, forming a
fundamental representation of this group.

The parameters of the potential of the Ginzburg-Landau theory
determine the ground state of the model; at high temperature, the
symmetry is unbroken, while at low temperature, either the AF or SC
order parameter attains an expectation value, depending on the doping.
Because of the coupling between the AF and SC order parameters, exotic
possibilities for solitons in the model can arise, as was observed
already by Zhang in his original paper. These ideas were developed in
Refs. \cite{abkz,abbg,golshe,maccli}.

In this paper, we wish to further analyze the properties of exotic
solitons in the SO(5) model.  We will consider in detail SC vortices
\cite{abkz,abbg}, although other possibilities \cite{golshe,maccli}
can be analyzed similarly. We will first introduce the SO(5) model and
review the reasons for suspecting that SC vortices might have AF
cores.

We will then study the free energy of vortices as a function of their
winding number.\footnote{This approach, implicit in the
  ground-breaking work of
  Bogomol'nyi \cite{bog}, is complementary to the usual
one of studying the surface energy density at a boundary between
normal and SC regions.}  Normally, in a type II superconductor (one
for which the Ginzburg-Landau parameter $\kappa$ satisfies
$\kappa>\kappa_c=1/\sqrt2$) the energy divided by winding number $m$
(or energy per flux quantum) of a vortex is an increasing function of
$m$. This implies that vortices of winding number greater than one are
unstable, which is one way of seeing that the vortex lattice is the
preferred (lowest-energy) configuration of a superconductor placed in
an external magnetic field. For a type I superconductor
($\kappa<\kappa_c$), the situation is reversed: the energy divided by
$m$ is a decreasing function of $m$ and the vortex lattice is
unstable.

However, as will be seen below, this is not necessarily true in the
SO(5) model.  Under certain circumstances, the vortex energy per flux
of a type II superconductor can be a {\em decreasing} function of
flux, indicating an instability of the vortex lattice: type I
behaviour.

The underlying reason is the possibility of an AF vortex core. When
this occurs, the AF order parameter makes a contribution to the vortex
free energy which is increasingly negative with increasing $m$.

Two factors are involved. The first is the degree to which a
superconductor is type II; the second is the proximity to SO(5)
symmetry, which is explicitly broken away from ``critical doping''
(that which corresponds to the SC-AF phase boundary). These factors
reinforce one another, so that a mildly type II superconductor can
easily exhibit type I behaviour, while a strongly type II
superconductor requires a doping exceedingly close to critical.

This feature of the SO(5) model gives, in principle, a dramatic
prediction of that model. If one varies the doping in a given
superconductor, the vortex lattice
should become less and less rigid, melting more and more easily as
critical doping is approached. Eventually, type I behaviour should
appear.

It must be noted that the region of parameter space corresponding to
critical doping appears to be experimentally delicate; in particular,
the appearance of inhomogeneities (stripe formation, phase separation)
\cite{whisca,ekt,vbbk} could mask the appearance of type I behaviour.
Nonetheless, reduced melting temperatures should appear away from this
delicate region, so that an experimental signature is still
possible. Indeed, Sonier, {\it et al.} have studied the melting of the
vortex lattice in high-temperature superconductors and have observed
melting at temperatures lower than expected in underdoped cuprates
\cite{sonier1}.

\medskip
\noindent{\bf 2. Vortices in the SO(5) Model.}
According to the SO(5) model, the low-energy dynamics of high
temperature superconductors is
written in terms of a 5-component real field transforming as a
fundamental representation of SO(5). The upper two components, say, of
this real field are the real and imaginary components of the complex
order parameter of superconductivity, while the lower three components are
the AF order parameter. We will call these fields
$\phi=\phi_1+i\phi_2$ and $\bfeta=(\eta_1,\eta_2,\eta_3)$,
respectively.

The low-energy effective theory can be described in terms of the
following free energy:
\bea
\hat F&=&\int d^2x\left({{\hat h}^2\over8\pi}+{\hbar^2\over2m^*}
\left|\left(-i\nabla-{e^*\over\hbar c}\hat\bfA\right)\phi\right|^2
\right.\nonumber\\
&&+ \left.{\hbar^2\over2m^*}(\nabla\bfeta)^2+V(\phi,\bfeta)\right),
\eea
where $\hat h=\nabla\times\hat\bfA$ is the microscopic magnetic field
(hats will simplify notation shortly, when we go to a description in
terms of dimensionless quantities).

Much information (including
the ground state) can be found by examining the potential. Including
even powers of the fields up to fourth order, the most general
potential is
\beq
V(\phi,\eta)=-{a_1^2\over2}\phi^2-{a_2^2\over2}\eta^2
+{b_1\phi^4+2b_3\phi^2\eta^2+b_2 \eta^4\over4}
\eeq
where we have written $\phi=|\phi|$ and $\eta=|\bfeta|$.
We have given the quadratic terms negative coefficients since this is what
is phenomenologically interesting.
In order for the potential to be
bounded from below, the quartic terms must obey the following
inequalities:
$b_{1,2}>0$, $b_3>-\sqrt{b_1b_2}$.

Strictly speaking, the model should
be called an SO(3)$\times$SO(2) model, since this is the
actual symmetry of the model. Nonetheless, the potential is invariant
under the larger group SO(5) if the two mass parameters are equal and
if the three quartic couplings are equal. It will be an approximate
symmetry if these couplings are approximately equal.
In what follows, for
simplicity we will set the three quartic couplings to the same value,
$b_1 = b_2 = b_3 =b$. 

In order to study SC vortices,
we must restrict ourselves to the region in parameter space that gives
a SC ground state.
This will be
the case if the global minimum of the potential has a nonzero value of
$\phi$ and a zero value of $\eta$. Examination of the potential shows
this to be true if $\be\equiv{a_2}^2/{a_1}^2<1$.
Then the
ground state is $(\phi,\eta)=(v,0)$, where $v=a_1/\sqrt{b}$. It is
convenient to add a constant $a_1^4/b$ to the potential, so
that the free energy of the superconducting state in the absence of a
magnetic field is zero.
Note that $\be=1$ corresponds
to the SO(5) symmetric limit of the potential, and also to critical
doping, since neither the SC or AF state
is preferred at that value.

It is easy to see qualitatively
why the core of a vortex {\em might\/}
have an AF core (i.e., a core where $\eta\neq0$).
In a vortex (in the SO(5) model as well as in the familiar
case of conventional
superconductors), the field $\phi$ changes in phase by $2\pi$ at
spatial infinity. By continuity, $\phi$ must have a zero at some point,
chosen to be the origin. Now let us look at how the field $\eta$ fits
into the situation.
At infinity, $|\phi|=v$ and the energy is minimized for
$\eta=0$. Inside the vortex core, however, $|\phi|\to0$. This
means that the potential, viewed as a function of $\eta$ with $\phi=0$,
is minimized at
$\eta\neq0$. Were the potential energy the only factor, $\eta$ would
certainly develop
a nonzero expectation value inside the core of the vortex. However
potential and gradient energy are in competition (the gradient energy
being minimized if $\eta$ is zero everywhere), and the minimum
energy configuration may or may not have $\eta\neq0$ in the core of the
vortex, depending on which of these two competing factors
dominates. The form of the potential suggests that as $\be$ is
increased, there is greater likelihood of an AF core; this is indeed
what is found numerically (see below, as well as Refs. \cite{abkz,maccli}).

As ansatz for the vortex, we use that of a conventional vortex
(generalized to winding number $m$) with in
addition an ansatz for $\eta$ (whose orientation is taken to
be constant) which allows for the possibility
of a nonzero core:
\bea
\phi(x)&=&v\,f(s) e^{i m\theta}\\
{\hat A}_i(x)&=&{a_1 c\sqrt{m^*}\over e^*}\epsilon_{ij}{s_j\over s}
A(s)\\
\eta(x)&=&v\,n(s)
\eea
where $s=r/\lambda$, $\lambda$ being the penetration depth,
$\lambda=(m^*c^2/4\pi{e^*}^2v^2)^{1/2}$. 

The equations of motion of the dimensionless fields $f,\ n$ and $A$
are (prime denotes derivative with respect to $s$):
\bea
{1\over\kappa^2}\left(f''+{1\over s}f'-\left({m\over s}
+\kappa A\right)^2\! f\right)+f(1\!-\!f^2\!-\!n^2)&=&0\\
{1\over\kappa^2}\left(n''+{1\over s}n'\right)+n(\be-f^2-n^2)
&=&0\\
h'+\left({m\over\kappa s}+A\right)f^2&=&0
\eea
where in the last equation $h$ is the dimensionless magnetic field,
defined by $h=-A'-A/s$.
The dimensionless free energy $F=(2e^{*2}/a_1^2m^*c^2)\hat F$
of a vortex of winding number $m$
is given by
\bea
F(m)&=&\int_0^\infty ds{s\over2}\Biggl\{
\left(A'+{A\over s}\right)^2 \nonumber\\
&+&\kappa^{-2}\left(f'^2+\left({m\over s}+\kappa A\right)^2f^2
+n'^2\right)\nonumber\\
&\qquad&\qquad\qquad
-f^2-\be n^2+\half\left(f^2+n^2\right)^2+\half\Biggr\}.
\label{dfa}
\eea
These expressions contain three parameters: the Ginzburg-Landau
parameter $\kappa=\lambda/\xi$ (where the coherence length is
$\xi=(\hbar^2/m^*{a_1}^2)^{1/2}$), the parameter $\be$ defined
above, and the
winding number of the vortex $m$. For high-temperature
superconductors, $\kappa$ is usually quite large, while $\be$ is
determined by sample preparation, by varying the doping.
(Specifically, $\beta$ can be written in
terms of more physical quantities as $\beta=1-8m^*\xi(T)^2
\chi(\mu_c^2-\mu^2)/\hbar^2$.)     $\be>1$
corresponds to the AF phase, while $\be<1$ describes the SC
phase. We will be particularly interested in $\be\ltwid1$.

\medskip
\noindent{\bf 3. Vortex energetics.}
For a given $m$, the vortex may or may not have an AF core, depending
on the parameters of the model. We define $\be_c(\kappa,m)$,
the critical value of $\be$, such that for $\be>\be_c$
the vortex core is AF, while for $\be<\be_c$ it is normal.
Figure 1 shows $\be_c$ as a function of
$\kappa$, for various values of $m$. One sees that as $m$ increases,
$\be_c$ decreases. This can be understood intuitively:
higher $m$ corresponds to a
wider vortex core, and thus greater impetus for $n$ to attain a
nonzero value in the core.

\begin{figure}[h]
\centerline{\psfig{file=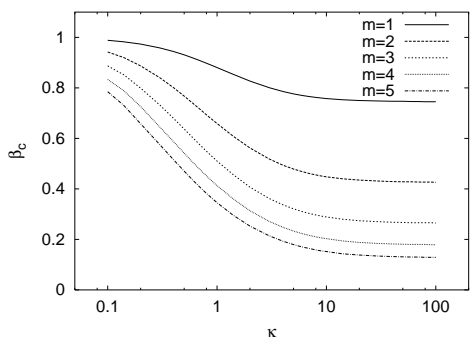,width=6.6cm,angle=-90}}
\caption{$\be_c$ as a function of $\kappa$ for various winding numbers.}
\end{figure}

A very useful quantity for given values of $\kappa$ and $\be$ is
the free energy per winding number of a vortex
as a function of $m$, $\FF(m)=F(m)/m$. This quantity clearly influences
the behaviour of a superconductor when placed in a magnetic
field: if $\FF$ increases with $m$, the field will penetrate in
vortices of winding number 1, while if $\FF$ decreases
with $m$, vortices will coalesce to form large normal regions.

For a conventional superconductor, $\FF$ increases or decreases with
$m$ for type-II or type-I superconductors, respectively.\footnote{Note
that we are defining the type of a superconductor according to the
value of $\kappa$ rather than according to the behaviour of the
superconductor in a magnetic field. Since for conventional
superconductors there is a simple relation between the two, this
distinction need not be made. In the SO(5) model, however, both
$\kappa$ and $\be$ influence the magnetic behaviour of the
superconductor.}  The SO(5) model gives $\FF(m)$ for a conventional
superconductor by setting $\be=0$; then, $n(s)=0$ and the 
vortex free energy
(\ref{dfa}) is identical to that of a conventional
superconductor.

Figure 2 shows $\FF(m)$ for various values of $\be$ and $\kappa$. In
the first three plots, the upper curve ($\be=0$) represents a conventional
superconductor: $\FF(m)$ is decreasing, constant and increasing for
type I, borderline I-II and type II superconductors, respectively. The
remaining curves reflect the effect of an AF core in the SO(5)
model. The fourth plot corresponds to a large value of $\kappa$;
$\be=0$ is not displayed in order to resolve different values of $\be$
very close to 1.

It is clear that the development of an AF core has a profound effect
on $\FF(m)$. This can be understood qualitatively in the following
way. As mentioned above, as $m$ increases, the vortex core width
increases. This is already true for conventional superconductors, but
the effect is more pronounced for SO(5) superconductors when the core
becomes AF, since in that case the free energy difference between the
AF and SC states is reduced, and the potential energy (which tends to
reduce the core size) is less important. Larger core size permits a
more spread out magnetic field, and an overall reduced energy.  (Note
that anomalously large core sizes in YBCO at low magnetic field have
been observed \cite{sonier2}, though whether the SO(5) model can
explain this has not yet been addressed.)

In a type I superconductor (Figure 2a)
$\FF(m)$ decreases more quickly once an AF
core develops. This changes in a quantitative way, but not a
qualitative way, the behaviour of the material.

Things are more interesting in the case of a type II superconductor
(Figures 2c, 2d),
where a qualitative transition from type II behaviour to type I can be
achieved. This occurs at approximately $\be=0.98$ and $\be=0.9998$ for
$\kappa=7.07$ and 70.7, respectively.

Clearly for strongly type II superconductors (as is the case with
high-temperature
superconductors), $\be$ must be extremely close to 1 (doping
extremely close to critical) for this transition to occur. Even before
this point, there is a substantial decrease in $\FF(m)$, meaning that
the energetic savings in forming a vortex lattice (as compared to a
large, normal region where the magnetic field pene-

\newpage

\twocolumn[\hsize\textwidth\columnwidth\hsize\csname@twocolumnfalse%
\endcsname
\centerline{{\psfig{file=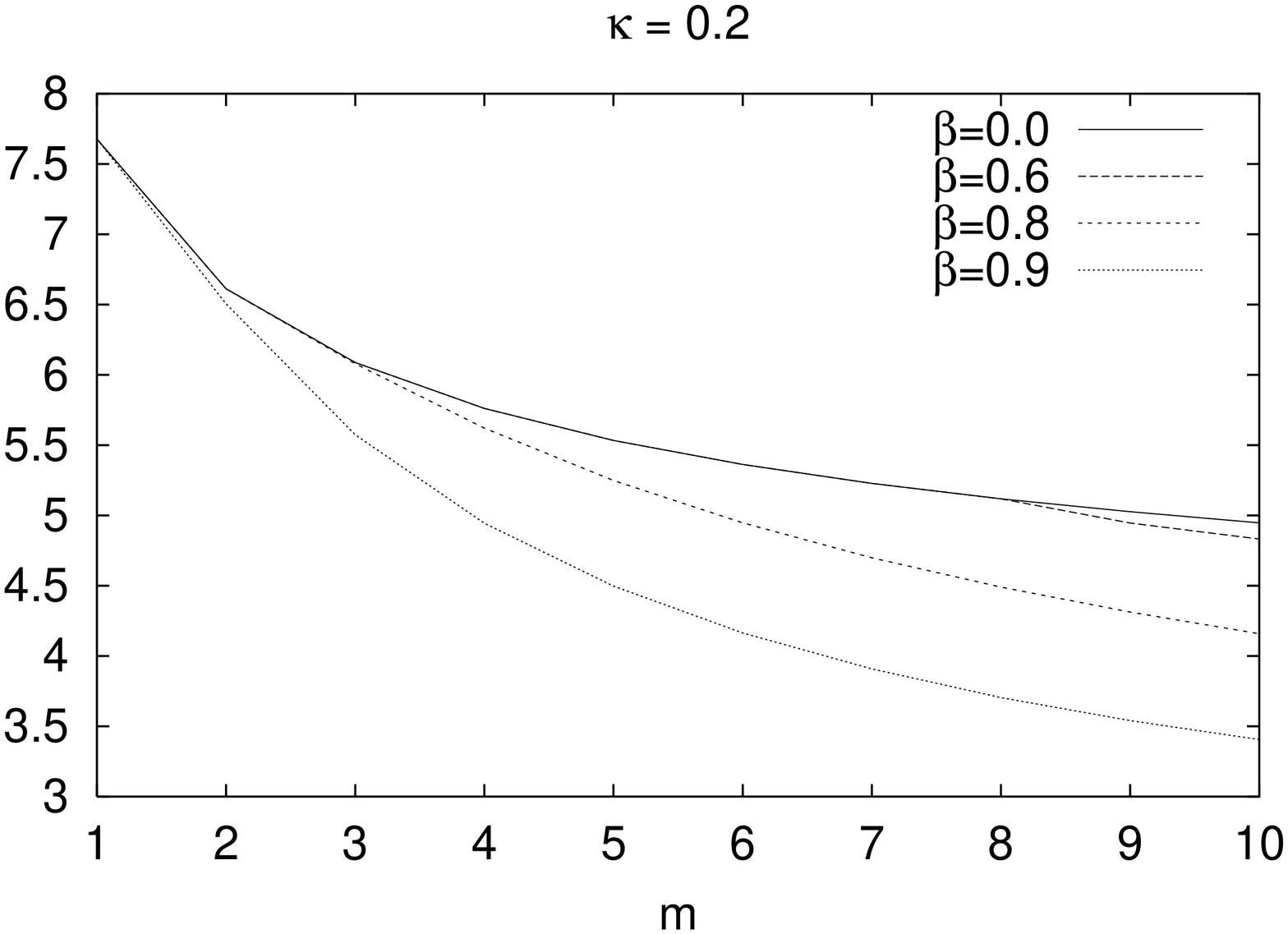,width=6.4cm,angle=-90}}
{\psfig{file=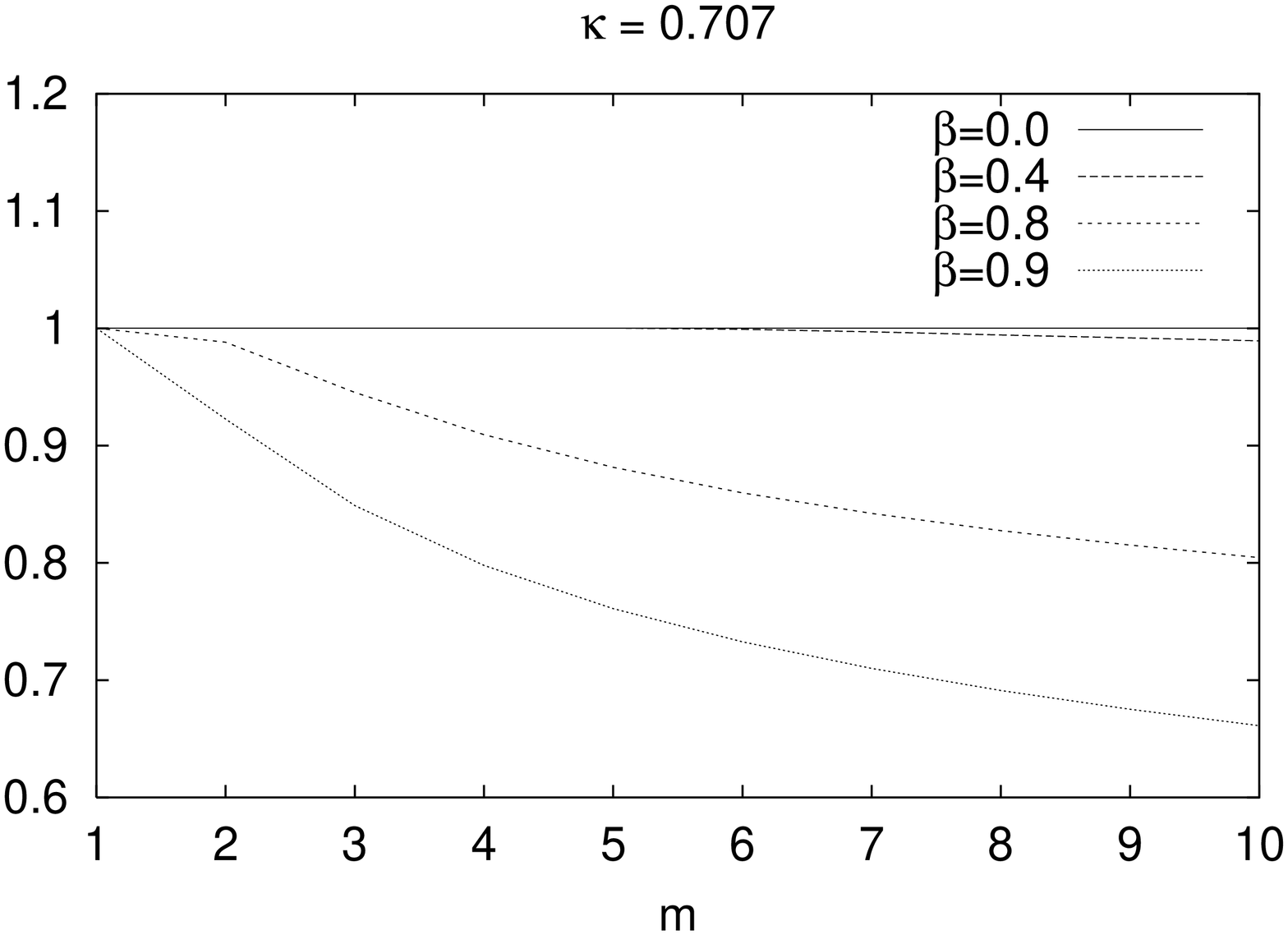,width=6.4cm,angle=-90}}}
\centerline{{\psfig{file=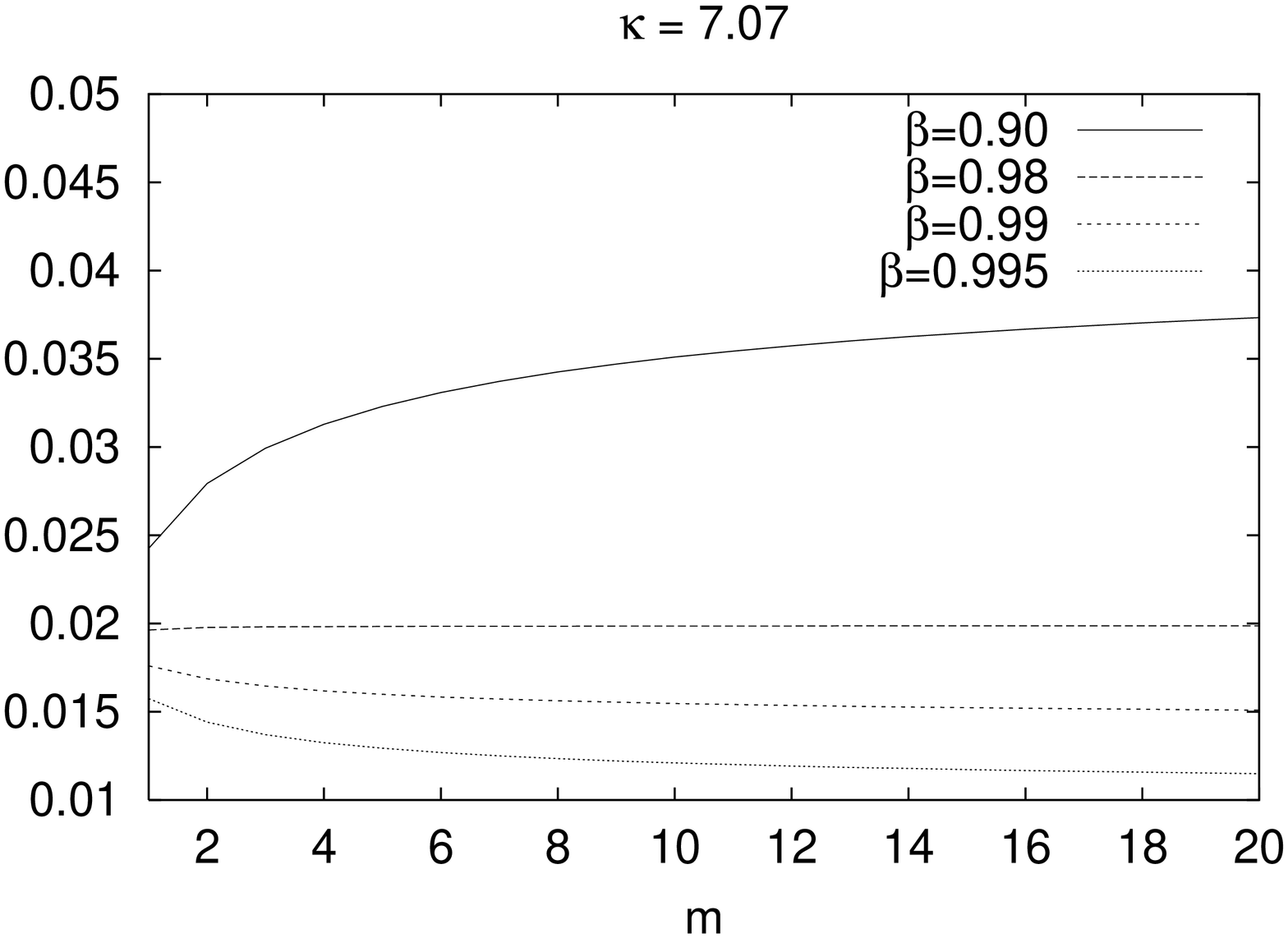,width=6.4cm,angle=-90}}
{\psfig{file=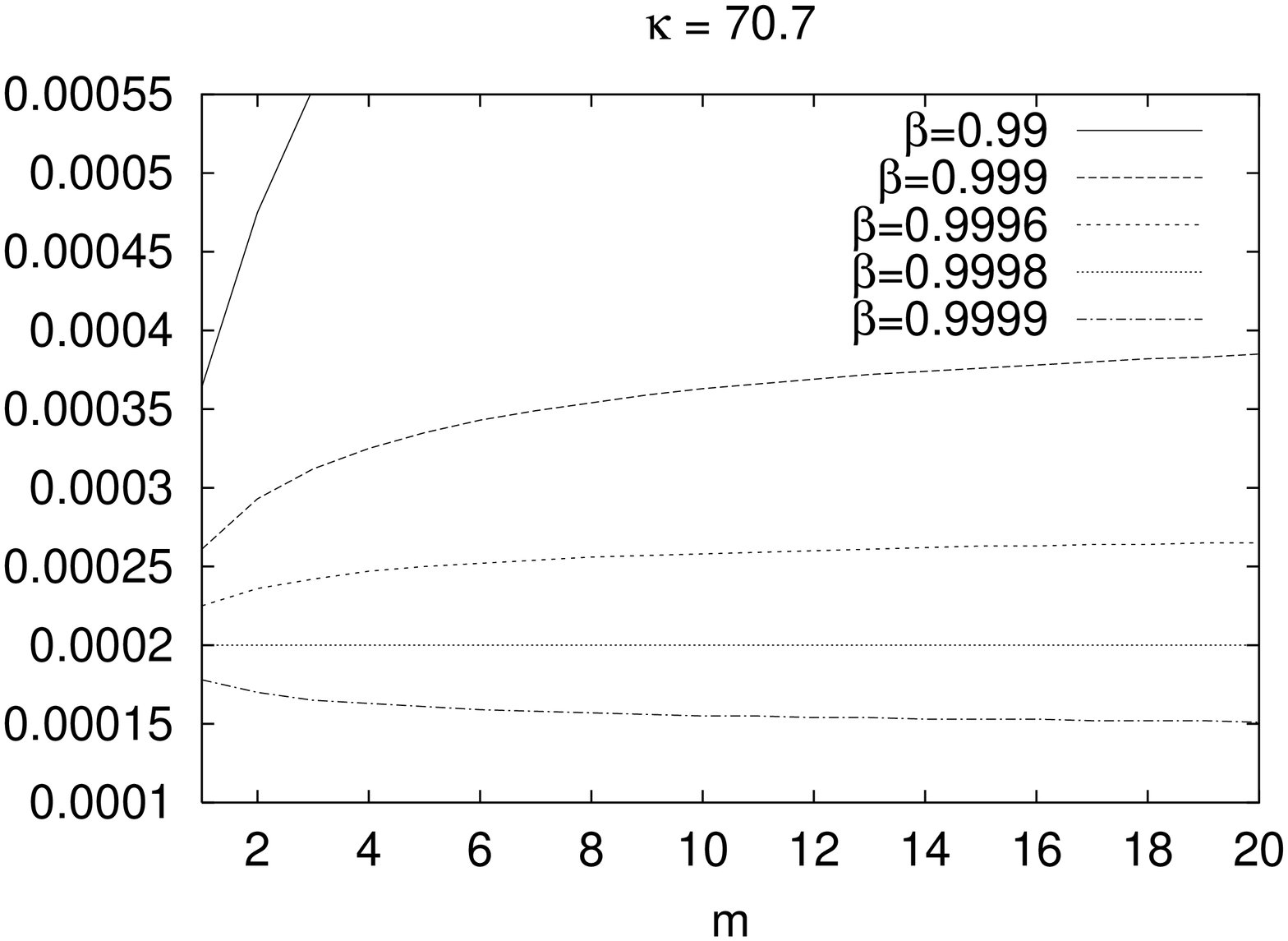,width=6.4cm,angle=-90}}}

\smallskip

\centerline{Figure 2:
$\FF(m)$ for various values of parameters $\kappa$ and $\be$.}
\medskip ]

\noindent trates) is
substantially reduced. This would be reflected in a less rigid, more
easily melted lattice.  Such behaviour has in fact been seen in
underdoped cuprates \cite{sonier1}.

We have also calculated the surface energy at a normal-superconducting
boundary as a function of $\be$ and $\kappa$, and find results
consistent with the above analysis: a positive or negative surface
energy when $\FF(m)$ is of negative or positive slope, respectively.
This will be reported elsewhere.

In summary, by analyzing the energy per unit flux of vortices as a
function of winding number in the SO(5) model,
we find that the development of an antiferromagnetic core has a
profound effect on the behaviour of a superconductor in a magnetic
field. This effect depends on the doping of the material, becoming
more and more strong as the doping is reduced to the critical value
(that corresponding to the AF/SC transition). More specifically, we find
that the degree to which a given superconductor behaves
as a type II superconductor decreases as the doping is reduced.
This can result in a less rigid (more easily melted) vortex lattice, and
as the doping approaches its critical value type I behaviour results.

Speight \cite{speight} has recently analyzed the static intervortex
force in conventional superconductivity, by treating the vortices as
point sources. It would be interesting to repeat this analysis in the
SO(5) model to see the effect of the $n$ field on these forces, and to
see if that the above behaviour can be understood in terms of
long-range forces between vortices.

It would also be interesting to extend the work of Bogomol'nyi
\cite{bog} to the
SO(5) model. This would circumvent much of the numerical work done in
the present article. We have not yet succeeded in doing so, however.

R.M.~thanks R. Kiefl for useful correspondence.
This work was supported in part by the Natural Science and Engineering
Research Council of Canada.

\end{document}